\documentclass{llncs}

\usepackage{graphicx}
\usepackage{subfigure}
\usepackage{amsmath}
\usepackage{color}

\title{Mutual Enrichment in Ranked Lists and the Statistical Assessment of Position Weight Matrix Motifs}
\titlerunning{Statistics in Ranked Lists and Position Weight Matrices}
\author{Limor Leibovich\inst{1} \and Zohar Yakhini\inst{1,2}}
\institute{Department of Computer Science, Technion -- Israel Institute of Technology, Technion City, Haifa 32000, Israel. \\ \email{llimor@cs.technion.ac.il} \and
Agilent Laboratories Israel, 94 Em Hamoshavot Road, 49527 Petach-Tikva, Israel. \email{zohar\_yakhini@agilent.com}}

\begin{document}

\maketitle

\begin{abstract}
 Statistics in ranked lists is important in analyzing molecular biology measurement data, such as ChIP-seq, which yields ranked lists of genomic sequences. State of the art methods study fixed motifs in ranked lists. More flexible models such as position weight matrix (PWM) motifs are not addressed in this context. To assess the enrichment of a PWM motif in a ranked list we use a PWM induced second ranking on the same set of elements. Possible orders of one ranked list relative to the other are modeled by permutations. Due to sample space complexity, it is difficult to characterize tail distributions in the group of permutations. In this paper we develop tight upper bounds on tail distributions of the size of the intersection of the top of two uniformly and independently drawn permutations and demonstrate advantages of this approach using our software implementation, mmHG-Finder, to study PWMs in several datasets.
\end{abstract}

\section{Introduction}

Modern data analysis often faces the task of extracting characteristic features from sets of elements singled out according to some measurement. In molecular biology, for example, an experiment may lead to measurement results pertaining to genes and then questions are asked about the properties of genes for which these were high or low. This is an example, of course, and the set of elements does not have to be genes. They can be genomic regions, proteins, structures, etc. A central technique for addressing the analysis of characteristic properties of sets of elements is statistical enrichment. More specifically -- the experiment results are often representable as ranked lists of elements and we then seek enrichment of other properties of these elements at the top or bottom of the ranked list. GSEA \cite{subramian-01}, for example, is a tool that addresses characteristic features of genes that are found to be differentially expressed in a comparative transcriptomics study. GOrilla \cite{eden-02} addresses GO 
terms enriched in ranked lists of genes where the ranking can be, for example, the result of processing differential expression data or of correlations computed between genomic DNA copy number and expression~\cite{ragle-03},\cite{akavia-04},\cite{dehan-05}. FATIGO \cite{shahorur-06} is also a tool that is useful in the context of analyzing GO terms in ranked lists of genes. DRIMust \cite{leibovich-07}, \cite{leibovich-08} searches for sequence motifs that are enriched, in a statistically significant manner, in the top of a ranked list of sequences, such as one produced by techniques like ChIP-seq.

All the aforementioned tools utilize a statistical approach that is based on assessing enrichment of an input set in an input ranked list by assessing the enrichment obtained at various cutoffs applied to the ranked list. It is often the case, however, that two quantitative properties need to be compared to each other. For example, when the elements are genes, we may have measured differential expression values, as well as measured ChIP-seq signals. We are therefore interested in assessing mutual enrichment in two ranked lists. Another example consists of one ranking according to differential expression and one according to prediction scores for miRNA targets. miTEA \cite{dsteinfeld-09} addresses this latter case by statistically assessing the enrichment of miRNA targets in a ranked list of genes (also see \cite{enerly-10}). To address mutual enrichment in two ranked lists over the same set of \textit{N} elements, miTEA \cite{dsteinfeld-09} performs analysis on permutations. Mutual enrichment in the top of 
two ranked lists can be simplified, from a 
mathematical point of view, by arbitrarily setting the indices of one list to the identity permutation \((1,2,...,\)\textit{N}) and treating the other list as a permutation \(\pi\) over these numbers. For the purpose of assessing the intersection of the top of the two ranked lists in a data driven manner, miTEA asks which prefix \([1,\dots,\textit{n}_1]\) is enriched in the first \(\textit{n}_2\) elements of that permutation,
\(\pi = \pi(1), ..., \pi(\textit{N})\). The statistics introduced by miTEA is called mmHG (min-min-Hyper-Geometric). A variant of mmHG is explained in detail in Section 2 of the current manuscript.%

Statistics in the group of permutations \(S_N\) is often difficult because the number of entities to be considered by any null model is \textit{N}! Direct exhaustive calculation of tail distributions over \(S_N\) is therefore impractical for all but very small values of \textit{N}. This difficulty is addressed by several heuristic techniques. Mapping into continuous spaces, such as in \cite{plis-11}, has proven useful in certain cases but not for studying large deviations. In the case of enrichment statistics that focuses on the top of the permutation and seeks to assess extreme events, such as mmHG, we prefer to use bounds on tail probabilities. Tail probabilities are useful constructs when applied to analyzing molecular biology measurement data as they enable statistical assessment of observed results. 

In this work we derive a tight bound on the tail probabilities of the mutual enrichment at the top of two random permutations uniformly drawn over \(S_N\) and demonstrate the utility of this approach in the context of flexible motif discovery. Our bounds are computable in polynomial time and potentially add to the accuracy of reported position weight matrix (PWM) motifs for nucleic acid sequences. 

\section{Background and Definitions}
\subsection{Mutual Enrichment in Ranked Lists -- the mmHG Statistics}

The mmHG statistics~\cite{dsteinfeld-09} is a generalization of the mHG statistics~\cite{eden-02},\cite{eden-12},\cite{steinfeld-13},
\cite{straussman-14}. While the mHG statistics quantifies the enrichment level of a set of elements at the top of a ranked list of elements of the same type, the mmHG statistics quantifies the level of mutual enrichment in two ranked lists over the same set of elements. While any parametric or non-parametric correlation statistics (e.g. Spearman's correlation coefficient), that takes the same input, calculates the overall agreement between the two ranked lists, the mmHG statistic focuses only on agreement at the top of the two ranked lists. mmHG counts elements common to the top of both lists, without predefining what top is. Its intended output is the probability for observing an intersection at least as large in two randomly ranked lists (the enrichment mmHG \textit{P-value}). In this section we describe the mmHG statistics and in later sections we suggest a tight bound for the p-value. Our definition of the mmHG statistic varies slightly from that of Steinfeld \textit{et al.}~\cite{dsteinfeld-09}.

Mutual enrichment in the top of two ranked lists can be simplified, from a mathematical point of view, by arbitrarily setting the indices of one list to the identity permutation \((1,2,...,\)\textit{N}) and treating the other list as a permutation. Details of this transform are given in Section 2.3. We now define mmHG for the simple case of one permutation. Consider a permutation \(\pi = \pi(1), ..., \pi(\textit{N}) \in S_N\) - 
the group of all permutations over the numbers \(1,..., \textit{N}\). mmHG is a function that takes \(\pi\) and calculates two numbers 
\(1 \leq n_1, n_2 \leq \textit{N}\) such that the observed intersection between the numbers \(1,...,n_1\) and the first \(n_2\) elements 
of \(\pi - \pi(1), ...,\pi(n_2)\) -- is the most surprising in terms of the hypergeometric p-value. Additionally, mmHG further calculates this aforementioned p-value.

Formally, given \(\pi \in S_N\) and for every \(1 \leq n_1, n_2 \leq \textit{N}\), let \(b_{\pi}(n_1,n_2)\) be the size of the intersection of \(1,...,n_1\) with \(\pi(1),...,\pi(n_2)\). Set \noindent
\begin{equation*} 
mmHG\ score(\pi)= \min_{1\leq n_1\leq N}\min_{1\leq n_2\leq N} \textsc{hgt}\left(N, n_1, n_2, b_{\pi}(n_1, n_2)\right)
\end{equation*}
where \textsc{hgt} is the tail distribution of an hypergeometric random variable:

\begin{equation*} 
\textsc{hgt}(N, n_1, n_2, b) = \sum\limits_{i=b}^{\min (n_1,n_2)} \frac{\binom{n_1}{i}\binom{N - n_1}{n_2 - i} }{\binom{N}{n_2}}
\end{equation*}

The mmHG score cannot be considered as a significance measure, due to the multiple testing involved in finding \(n_1\) and \(n_2\). 
A simple way to correct an mmHG score s for multiple testing and report a p-value bound would be to use the Bonferroni correction. That is done by multiplying s by the number of multiple tests conducted which is \(N^{2}\). Therefore:
\begin{equation*}
mmHG\ p -value(s,N) \leq s \cdot N^2 
\end{equation*}
In Section 3 we present significantly tighter bounds.

\subsection{PWM Motifs}
Data produced by techniques such as ChIP-seq~\cite{lee-15}, ChIP-exo~\cite{rhee-16}, CLIP~\cite{lebedeva-17}, PAR-CLIP~\cite{hafner-18} and others are readily representable as ranked lists of sequences, where the ranking is according to measured binding affinity. Computational tools and approaches to motif discovery form part of the data analysis workflow that is used to extract knowledge and understanding from this type of studies.
We are often interested in sequence motifs that are observed to be enriched in sequences where strong binding affinity is measured. A position weight matrix (PWM) is a commonly used representation of motifs in biological sequences~\cite{staden-19},\cite{stormo-20},\cite{hertz-21}. This representation is more faithful to biology than representation by exact words. A PWM is a matrix of score values that gives a weighted match to any given substring of fixed length. It has one row for each symbol of the alphabet, and one column for each position in the pattern. The score assigned by
a PWM to a\ substring \textit{S} = \(\textit{S}_1...\textit{S}_K\) is defined as \(\sum_{j=1}^{K} m_{s_{j},j}\), where 
\textit{j} represents position in the substring, \(\textit{S}_j\)\ is the symbol at position~\textit{j} ~in the substring, and 
\(m_{\alpha,j}\) is the score in row ~\(\alpha\), column~\textit{j} of the matrix. In other words, a PWM score is the sum of position-specific scores for each symbol in the substring. This definition can be generalized to yield a score for a sequence 
\textit{S} = \(\textit{S}_1...\textit{S}_M\) longer than the PWM by calculating 
$max_{1\leq i\leq M-K+1}\sum_{j=1}^{K} m_{s_{i+j-1},j}$. Alternatively, an enhanced model that takes into account multiple occurrences of the PWM in the sequence can be applied by summing over sufficiently strong occurrences of the PWM or by other  more sophisticated approaches~\cite{dsinha-22}.

\subsection{mmHG Statistics for PWM Motifs}

Given a set of sequences that were tested in a high throughput experiment such as ChIP-seq~\cite{lee-15}, CLIP~\cite{lebedeva-17} and others, they can be ranked according to the measured binding affinities, yielding a ranked list \(L_1\). Since usually we are interested in finding motifs amongst sequences having strong binding affinities, we actually search for motifs that are more prevalent at the top of this list. It is clear that any algorithm for de-novo flexible motif search would need to evaluate candidate PWMs. Given a PWM which we want to assess, the sequences can also be ranked according to their PWM scores, yielding another ranked list \(L_2\), different from \(L_1\). A significant PWM motif would yield significant scores for sequences having strong binding affinities. Therefore, the question of PWM motif discovery from ranked experimental data can be formulated as quantifying the mutual enrichment level for the two ranked lists \(L_1\) and \(L_2\). Given two ranked lists \(L_1\) and 
\(L_2\) over the universe of \textit{N} sequences, they can be transformed into two respective permutations, 
\(\pi_1 = (\pi_1(1),...,\pi_1(N))\) and \(\pi_2 = (\pi_2(1),...,\pi_2(N))\). The relative permutation \(\pi\), of \(\pi_2\) w.r.t. \(\pi_1\), is defined by \(\pi(\pi_1(j)) = \pi_2(j)\), for every \(\textit{j} = 1,...,\textit{N}\) or simply, using the operations in the group 
\(S_N: \pi = \pi_2 \cdot \pi_{1}^{-1}\). Using the relative permutation \(\pi\), we can represent the mutual enrichment of the top parts of 
\(L_1\) and \(L_2\) as \textit{mmHG} \textit{score}(\(\pi\)), defined above.

\section{Algorithms and Results}

\subsection{Estimation of the mmHG p-value -- Introducing First Upper Bound}

Given an mmHG score \textit{s}, observed in analyzing real measurement data, we would like to assess the statistical significance of this observation. Assuming endless computational power, we would enumerate all permutations and calculate the mmHG score for each, in order to characterize the distribution of mmHG as a random variable over \(S_N\). The p-value for \textit{s} is then simply:
\begin{equation*}
mmHG\ p-value(s,N) = \frac{\text{\small The number of permutations having mmHG score} \leq s}{N!}
\end{equation*}

Since the number of permutations is huge, the process described above is very far from feasible. Therefore, we seek a computationally tractable upper bound, preferably tight.

A trivial upper bound is the Bonferroni corrected mmHG score defined by \(s\cdot N^2\). A more subtle upper bound was suggested by Steinfeld \textit{et al.}~\cite{dsteinfeld-09} and is briefly described in Section 3.3. In this work we introduce a tighter bound that is polynomially computable.

We will next describe an intuitive upper bound and later refine it to produce a tighter bound. Our input comprises an mmHG score \textit{s}, and the total number of elements \textit{N}. The output will be an upper bound for the p-value. The efficiency of our approach relies on enumerating all possible \textsc{hgt} scores rather than enumerating all permutations in \(S_N\). This approach is computationally efficient as \textsc{hgt} is a function of four input parameters: \textit{N}, \(\textit{n}_1\), \(\textit{n}_2\), and \textit{b}. Given \textit{N}, there are \(O(\textit{N}^3)\) possible combinations of \(\textit{n}_1\), \(\textit{n}_2\) and \textit{b}. Next, all is left to do is to determine how many permutations stand behind each \textsc{hgt} score. To this end, we will define the function $\Lambda(\textit{N},\textit{n}_1,\textit{n}_2,\textit{b})$ to be the number of permutations for which it holds that out of the first \(n_2\) entries, \textit{b} of them are taken from the range \([1,\dots,n_1]\). This 
formulation is equivalent to
counting permutations for which we attain, 
at some point, the value \textsc{hgt}\((\textit{N},\textit{n}_1,\textit{n}_2,\textit{b})\), had we taken the exhaustive approach.
\(\Lambda(\textit{N},\textit{n}_1,\textit{n}_2,\textit{b})\) can be represented as:
\begin{equation*}
\Lambda(\textit{N},\textit{n}_1,\textit{n}_2,\textit{b}) = \binom{n_1}{b}\binom{n_2}{b}b!\binom{N- n_1}{n_2 - b}(n_2 - b)!(N - n_2)!
\end{equation*}
as we first choose \textit{b} elements from the range \([1,\dots,n_1]\) to appear at the first \(\textit{n}_2\) entries of the permutation (there are 
\(\binom{n_1}{b}\) possibilities). Then, we choose where to position these \textit{b} elements at the first \(\textit{n}_2\) entries of the permutation and consider all internal arrangements (for each choice of \textit{b} elements there are \(\binom{n_2}{b}b!\) possibilities). We next choose \(\textit{n}_2-\textit{b}\) elements from the range \([\textit{n}_{1}+1,\dots,\textit{N}]\) to appear at the rest of the first \(\textit{n}_2\) entries of the permutation (there are \(\binom{N - n_1}{n_2 - b}\) possibilities for that) and consider all possible \((n_{2}-b)!\) arrangements. Finally, we take into account all possible \((\textit{N} - \textit{n}_2)!\) arrangements of the rest \textit{N}-\(\textit{n}_2\) entries of the permutation.

A straightforward upper bound for the number of permutations in \(S_N\) having mmHG score better than \textit{s} follows: 
\begin{align*}
\left|\left\{\pi' \in S_N: mmHG(\pi') \leq s\right\}\right| &\leq \sum_{\substack{n_1,n_2,b:\\\textsc{hgt}(N,n_1,n_2,b)\leq s}} \Lambda(N,n_1,n_2,b)
\intertext{From which an upper bound is easily derived:}
mmHG\ p - value(s,N) &\leq \frac{\sum_{\substack{n_1,n_2,b:\\\textsc{hgt}(N,n_1,n_2,b)\leq s}} \Lambda(N,n_1,n_2,b)}{N!}
\intertext{By algebraic manipulations we get:}
 mmHG\ p - value(s,N) &\leq \sum_{\substack{n_1,n_2,b:\\\textsc{hgt}(N,n_1,n_2,b)\leq s}} \frac{\binom{n_1}{b}\binom{N - n_1}{n_2 - b}}{\binom{N}{n_2}}
\end{align*}
This upper bound is simple and requires O\((\textit{N}^3)\) \textsc{hgt} calculations. An \textsc{hgt} calculation takes O(\textit{N}) time, assuming binomial coefficients can be calculated in O(1) time, for example by using Stirling's approximation~\cite{abramowitz-23}:\\
\(\sqrt{2\pi n}(\frac{n}{e})^n \frac{1}{e^{12n+1}}\leq n! \leq \sqrt{2\pi n}(\frac{n}{e})^n \frac{1}{e^{12n}}\) , which is tight for large factorials.

\subsection{A Refined Upper Bound for the p-value}

The upper bound introduced in the previous section counts the number of permutations for which the value \textsc{hgt}\((\textit{N},\textit{n}_1,\textit{n}_2,\textit{b})\) is calculated when taking the non-practical exhaustive approach that enumerates over all \textit{N}! permutations. Ideally, we wish to count the number of permutations for which the value \textsc{hgt}\((\textit{N},\textit{n}_1,\textit{n}_2,\textit{b})\) is also their mmHG score, as a permutation may have several \textsc{hgt} values that are better than \textit{s}, so it can be counted more than once. This explains why the formula introduced earlier is an upper bound and not an exact p-value. A second observation that follows is that the smaller the mmHG score \textit{s} is, the tighter the bound, because a permutation will have fewer combinations \((\textit{N},\textit{n}_1,\textit{n}_2,\textit{b})\) having \textsc{hgt} score better than \textit{s}.

Therefore, if we can reduce the extent of multiple counting of the same permutation, we will get a tighter bound. We do this by looking one step backwards. If, for example, \textsc{hgt}\((\textit{N},\textit{n}_1,\textit{n}_2,\textit{b}) \leq s\), we can exclude from the counting permutations that contain \textit{b} elements from the range \([1,\dots,\textit{n}_{1}-1]\) at their first \(\textit{n}_2\) entries because they are already taken into account in \(\Lambda(\textit{N},\textit{n}_{1} - 1,\textit{n}_2,\textit{b})\) (because necessarily \textsc{hgt}\((\textit{N},\textit{n}_{1} - 1,\textit{n}_2,\textit{b}) \leq s\), as we will later explain).

Let \(\Psi(\textit{N},\textit{n}_1,\textit{n}_2,\textit{b})\) be the set of permutations for which it holds that out of the first \(\textit{n}_2\) entries, \textit{b} of them are taken from the range \([1,\dots,\textit{n}_1]\) (note that \(\Lambda(\textit{N},\textit{n}_1,\textit{n}_2,\textit{b})\) introduced earlier is, therefore, the size of \(\Psi(\textit{N},\textit{n}_1,\textit{n}_2,\textit{b})\)). Assuming \textsc{hgt}\((\textit{N},\textit{n}_1,\textit{n}_2,\textit{b}) \leq s\), we can partition the set \(\Psi(\textit{N},\textit{n}_1,\textit{n}_2,\textit{b})\) into five disjoint subsets \(\psi_1,...,\psi_5\) such that
\(\psi = \psi_1 \cup \psi_2 \cup\psi_3 \cup\psi_4 \cup \psi_5\), as follows:
\begin{align*}
\psi_1 =\ &\Psi(\textit{N},\textit{n}_1,\textit{n}_2,\textit{b}) \cap \Psi(\textit{N},\textit{n}_{1} -1,\textit{n}_2 - 1,\textit{b} - 1) \cap
\Psi(\textit{N},\textit{n}_1 - 1,\textit{n}_2,\textit{b}) \\
\psi_2 =\ &\Psi(\textit{N},\textit{n}_1,\textit{n}_2,\textit{b}) \cap \Psi(\textit{N},\textit{n}_{1} -1,\textit{n}_2 - 1,\textit{b} - 1) \cap
\Psi(\textit{N},\textit{n}_1,\textit{n}_2 - 1,\textit{b}) \\
\psi_3 =\ &\Psi(\textit{N},\textit{n}_1,\textit{n}_2,\textit{b}) \cap \Psi(\textit{N},\textit{n}_{1} -1,\textit{n}_2 - 1,\textit{b} - 1) \cap
		\Psi(\textit{N},\textit{n}_1 - 1,\textit{n}_2,\textit{b} - 1)\\
	&\cap \Psi(\textit{N},\textit{n}_1,\textit{n}_2 - 1,\textit{b} - 1) \\
\psi_4 =\ &\Psi(\textit{N},\textit{n}_1,\textit{n}_2,\textit{b}) \cap \Psi(\textit{N},\textit{n}_{1} -1,\textit{n}_2 - 1,\textit{b}) \\
\psi_5 =\ &\Psi(\textit{N},\textit{n}_1,\textit{n}_2,\textit{b}) \cap \Psi(\textit{N},\textit{n}_{1} -1,\textit{n}_2 - 1,\textit{b} - 2) \cap
\Psi(\textit{N},\textit{n}_1 - 1,\textit{n}_2,\textit{b} - 1) \\
	&\cap \Psi(\textit{N},\textit{n}_1,\textit{n}_2 - 1,\textit{b} - 1)
\end{align*}

The properties of the hypergeometric distribution imply that \(\psi_1,\psi_2,\psi_4\) can be disregarded, in the current counting stage. To explain why, we will demonstrate the argument on \(\psi_1\). The permutations in \(\psi_1\) contain \textit{b} elements from the 
range \([1,\dots,\textit{n}_1 - 1]\) at the first \(\textit{n}_2\) entries. We also assume that 
\textsc{hgt}$(\textit{N},\textit{n}_1,\textit{n}_2,\textit{b})\!\leq s$. Therefore \textsc{hgt}\((\textit{N},\textit{n}_1 - 1,\textit{n}_2,\textit{b}) \leq s\) also holds, as the same intersection is observed for even a smaller set. Thus, the permutations in \(\psi_1\) should have been counted when handling the triplet \(\textit{n}_1-1\), \(\textit{n}_2\) and \textit{b} and disregarded for the combination \(\textit{n}_1\), \(\textit{n}_2\) and \textit{b}. Similar arguments hold for \(\psi_2\) and \(\psi_4\).

 \(\psi_3\) should be counted if it holds that \textsc{hgt}\((\textit{N},\textit{n}_1 - 1,\textit{n}_2 - 1,\textit{b} -1) > s\) and
\textsc{hgt}\((\textit{N},\textit{n}_1 - 1,\textit{n}_2,\textit{b} -1) > s\) and \textsc{hgt}\((\textit{N},\textit{n}_1,\textit{n}_2 - 1,\textit{b} - 1) > s\), otherwise \(\psi_3\) would have been counted by former triplets. Similarly, \(\psi_5\) should be counted if
\textsc{hgt}\((\textit{N},\textit{n}_1 - 1,\textit{n}_2 - 1,\textit{b} - 2) > s\) and \textsc{hgt}\((\textit{N},\textit{n}_1 - 1,\textit{n}_2,\textit{b} - 1) > s\) and \textsc{hgt}\((\textit{N},\textit{n}_1,\textit{n}_2 -1,\textit{b} - 1) > s\). Finally, we calculate the sizes of \(\psi_3\) and \(\psi_5\), in the relevant cases. The permutations in \(\psi_3\) contain \textit{b}-1 elements taken from the range \([1,\dots,\textit{n}_1 - 1]\) located at the first \(\textit{n}_2 - 1\) entries, where the number \(\textit{n}_1\) is positioned at entry \(\textit{n}_2\). Therefore:
\begin{equation*}
|\psi_3| = \binom{n_1 - 1}{b - 1}\binom{n_2 - 1}{b - 1}(b - 1)!\binom{N - n_1}{n_2 - b}(n_2 -b)!(N - n_2)!
\end{equation*}
The permutations in \(\psi_5\) contain \textit{b}-2 elements taken from \([1,\dots,\textit{n}_1 - 1]\) located at the first 
\(\textit{n}_2 - 1\), where \(\textit{n}_1\) is positioned at one of the first \(\textit{n}_2 - 1\) entries, and also entry \(\textit{n}_2\) 
contains an element from \([1,\dots,\textit{n}_1 - 1]\). Therefore:
\begin{equation*}
|\psi_5| = 
\binom{n_1 - 1}{b - 2}\binom{n_2 - 1}{b - 2}(b - 2)!(n_2 - b + 1)\binom{N - n_1}{n_2 - b}(n_2 -b)!(n_1 - b + 1)(N - n_2)!
\end{equation*}
From the above we next conclude an upper bound. Denote
\begin{equation*}
    I(\textsc{hgt}(\textit{N},\textit{n}_1,\textit{n}_2,\textit{b}) > s)=\left\{\begin{array}{ll} 1, & \ \text{if}\ \textsc{hgt}(\textit{N},\textit{n}_1,\textit{n}_2,\textit{b}) > s \\
         0, & \ \text{otherwise}\end{array}\right. .
\end{equation*}

\begin{align*}
\Lambda^{*}(\textit{N},\textit{n}_1,\textit{n}_2,\textit{b}) = \hspace{2em}&\\
|\psi_3| &\times I(\textsc{hgt}(\textit{N},\textit{n}_1 - 1,\textit{n}_2 - 1,\textit{b} - 1) > s) \\
         &\times I(\textsc{hgt}(\textit{N},\textit{n}_1 - 1,\textit{n}_2,\textit{b} - 1) > s) \\
         &\times I(\textsc{hgt}(\textit{N},\textit{n}_1,\textit{n}_2 - 1,\textit{b} - 1) > s) \\
&+ \\
|\psi_5| &\times I(\textsc{hgt}(\textit{N},\textit{n}_1 - 1,\textit{n}_2 - 1,\textit{b} - 2) > s) \\
         &\times I(\textsc{hgt}(\textit{N},\textit{n}_1 - 1,\textit{n}_2,\textit{b} - 1) > s) \\
         &\times I(\textsc{hgt}(\textit{N},\textit{n}_1,\textit{n}_2 - 1,\textit{b} - 1) > s) \\
\end{align*}%
Yielding the following upper bound for the p-value:
\begin{equation*}
mmHG\ p -value(s,N) \leq \frac{\sum_{\substack{n_1,n_2,b:\\\textsc{hgt}(N,n_1,n_2,b)\leq s}} \Lambda^{*}(N,n_1,n_2,b)}{N!}
\end{equation*}
Note that when \(n_1\) or \(n_2 \leq 1\), \(\Lambda^{*}(\textit{N},\textit{n}_1,\textit{n}_2,\textit{b})\) is defined as 
\(\Lambda(\textit{N},\textit{n}_1,\textit{n}_2,\textit{b})\). Also, given \textit{N}, \(\textit{n}_1\) and \(\textit{n}_2\), \textit{b} can be any integer in \([max(0,n_2 -N+n_1), min(n_1,n_2)]\).
 
 This upper bound uses more delicate counting than the bound introduced in the previous section. In the following sections we assess the tightness of this bound. In later sections we demonstrate an application for PWM motif search.

\subsection{Comparison to a Different Variant}

We note that the bound described in Steinfeld \textit{et al.}~\cite{dsteinfeld-09} addresses a slightly different variant of mmHG as a random variable over \(S_N\). The definition with which we work here is more amenable to deriving tight bounds as described above. Given a single permutation \(\pi \in S_N\) and for every \(\textit{i}=1,\dots,\textit{N}\), a binary vector \(\lambda_i\) is defined in which exactly \textit{i} entries are 1 and \textit{N}-\textit{i} entries are 0, as follows: \(\lambda_i(j) = 1\) if \(\pi(j) \leq i\). The mmHG score of a permutation 
\(\pi\) is then defined by Steinfeld \textit{et al}.~\cite{dsteinfeld-09} as:
\begin{equation*}
mmHG(\pi) = \min_{1\leq i\leq N} P - value(mHG(\lambda_i)),
\end{equation*}
where mHG($\lambda$) = $ \min_{1\leq i\leq N} \textsc{hgt}(\textit{N},B,n,\textit{b}_n), N = |\lambda|, b_n = \sum_{i=1}^{n} \lambda_i$
\text{and} $B = b_N.$
A possible upper bound is then given by:
\begin{equation*}
(*)\ \  P - value(mmHG(\pi)) \leq \min_{1\leq i\leq N} mHG(\lambda_i)\cdot i \cdot N
\end{equation*}
Computing the latter quantity requires O(\(\textit{N}^2\)) \textsc{hgt} calculations and is therefore more computationally efficient than the two bounds described in Sections 3.1 and 3.2 of this current work, that require O(\(\textit{N}^3\)) \textsc{hgt} calculations. We observed that our bound was tighter than the bound in (*), as later shown in Figure 1D. For example, for a permutation having mmHG score = \(7.8 \cdot 10^{-25} (\textit{N}=100)\), our bound was \(3.5\cdot10^{-23}\) while (*) yielded \(4.2\cdot 10^{-21}\). For one permutation with mmHG score = \(5.1\cdot 10^{-5} (\textit{N}=100)\),
our bound was 0.026 while (*) yielded 0.2. The latter example demonstrates that a tighter bound is important for classifying an observation as statistically significant (assuming a significance threshold of 0.05).

\subsection{Assessment of Tightness}
 
In order to assess the quality of our bound, we compared it to the exact p-value, which can be calculated for small values of \textit{N} (that is, in cases where \textit{N}! is not too large). Figure 1A compares the mmHG score (which also serves as a lower bound for the p-value), the exact p-value (calculated by exhaustive enumeration of all 10! permutations), our upper bound and the Bonferroni corrected p-value for \textit{N}=10. Figure 1B shows the same for \textit{N}=20, except that exact p-values cannot be calculated exhaustively, and therefore an empirical p-value is produced by randomly sampling \(10^{7}\) permutations. In both cases our upper bound is significantly tighter than the Bonferroni bound. We also observed that the smaller the mmHG scores are -- the tighter is our bound, consistent with lesser over-counting for smaller scores, as explained in previous sections. Comparison between the first bound described in Section 3.1 and the bound described in Section 3.2 is shown in Figure 1C (for \
textit{
N}=20). We observed that enumerating all \textsc{hgt} scores rather than enumerating all permutations in \(S_N\) significantly improves the p-value estimation. Moreover, the refinement of this approach produced by reducing the extent of multiple counting of permutation further improves the upper bound. In Figure 1D the bounds, including the bound introduced in Section 3.3 (Steinfeld bound), are shown for \textit{N}=100. An empirical p-value was not calculated here as even if we sample \(10^{7}\) permutations, a p-value smaller than \(10^{-7}\) cannot be obtained. The bound suggested in this paper was almost always observed to be tighter than the bound introduced in Section 3.3.

\begin{figure}[!ht]%
\centering%
\subfigure[]{\includegraphics[scale=.8]{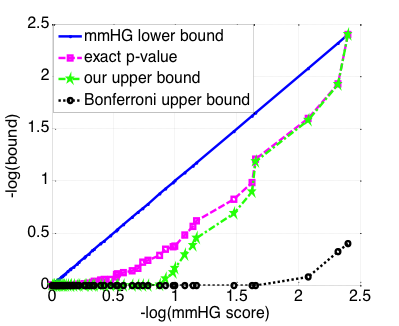}}%
\subfigure[]{\includegraphics[scale=.8]{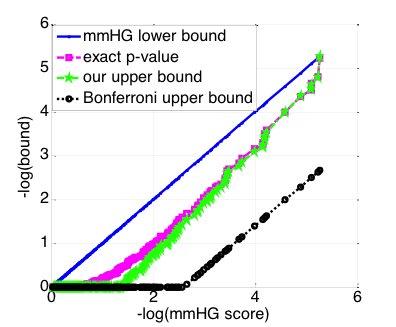}}
\subfigure[]{\includegraphics[scale=.8]{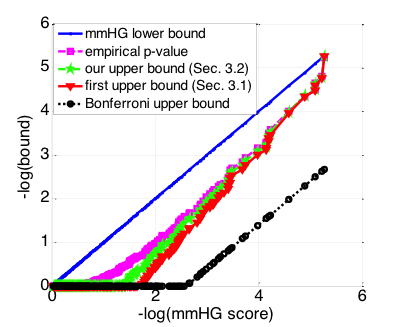}}%
\subfigure[]{\includegraphics[scale=.8]{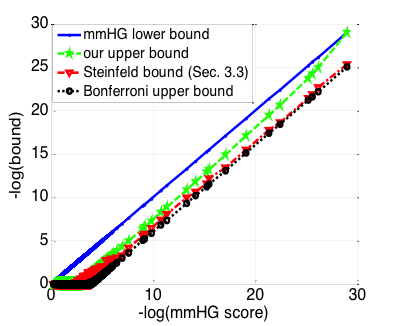}}%
\caption{(A) Four lines are shown for \textit{N}=10: the mmHG score, which also serves as a lower bound for the p-value; the exact p-value calculated by enumerating all 10! permutations; our upper bound, in its refined version; and the Bonferroni corrected p-value. (B) Here again the four lines are shown -- for \textit{N}=20. However, instead of an exact p-value, which cannot be calculated exhaustively, an empirical p-value is produced by randomly sampling \(10^7\) permutations. (C) In addition to the four lines shown in B, the first upper bound -- introduced in Section 3.1 - is shown (\textit{N}=20). (D) Four lines are shown for \textit{N}=100: the mmHG score, our upper bound, the bound introduced in Section 3.3 (Steinfeld bound) and the Bonferroni corrected p-value. The exact p-value line is positioned between the green and the blue lines.}%
\end{figure}

\subsection{Application in PWM Motif Search}

In this section we discuss mmHG as a framework for assessing the significance of PWM motifs in ranked lists. Given a ranked list of sequences and a PWM motif, by using the mmHG statistics and the bounds introduced earlier, we can assign a p-value to represent the significance of that PWM being enriched at the top of the list. To apply this approach for de-novo motif search, one needs to theoretically consider all possible PWMs. This is not feasible and as a heuristic approach we wrote mmHG-Finder which takes as input a ranked list of DNA or RNA sequences and returns significant motifs in PWM format. In cases where sequence ranking is not relevant or not available, it allows the use of positive and negative sets of sequences, searching for enriched motifs in the positive set using the negative set as the background.\medskip

We will now describe the methodology implemented in mmHG-Finder: 

\noindent
\underline{Input:}
\begin{itemize}
\renewcommand{\labelitemi}{$\bullet$}
\item a ranked list of sequences (or, alternatively, two sets of sequences representing target and background)
\item motif width, given as a range between \(k_1\) and \(k_2\)
\end{itemize}

\noindent
\underline{Algorithm:} 
\begin{enumerate}\small
 \item Build a generalized suffix tree for the sequences
\item  Traverse the tree to find all k-mers for \textit{k}=\(k_1,\dots,k_2\)
\item Sort the \textit{k}-mers according to their enrichment at the top of the list (this is done using the mHG statistics), as explained in 
Leibovich \textit{et al}.~\cite{leibovich-07}
\item Take the most significant fifty \textit{k}-mers, to be used as starting points for the next step. This set of candidates is chosen such that the members are quite different. Note that this is a heuristic approach and the number 50 is somewhat arbitrary, chosen to succeed in catching the best performing PWMs without heavily paying in complexity.
\item For each starting point, we iteratively replace one position in the \textit{k}-mers by considering all possible IUPAC replacements and taking the one that improves the enrichment the most. We repeat this process for all positions several times. Eventually we get a motif in the IUPAC alphabet which is then converted to a PWM.
\item The PWMs found in the previous step are assessed using the mmHG statistics and the best is returned as output, together with the p-value. The score assigned by a PWM to a string $S = S_1,\dots,S_M$ is defined as
$max_{1 \leq i\leq M-K+1}\sum_{j=1}^{K} m_{S_{i+j-1},j}$
(assuming \(M \geq K\), otherwise it is \(-\infty\)), where \(m_{\alpha,c}\) is the score in row \(\alpha\), column c of the position weight matrix. In other words, the PWM score calculated for S is the maximal score obtained for a substring of S.
\end{enumerate}
To evaluate the performance of mmHG-Finder in comparison to other state-of-the-art methods we ran it on 18 example cases -- 3 synthetically generated cases and 15 generated from high throughput binding experiments (6 transcription factors and 9 RNA-binding proteins). We compared the results to those obtained by using three other methods: the standard MEME program~\cite{bailey-24}, DREME~\cite{bailey-25}, and XXmotif~\cite{luehr-26}. Some of the results of this comparison are summarized in Table 1. The synthetic examples were generated by randomizing 500 sequences of length 100. An IUPAC motif was generated and planted in all top 64 sequences. mmHG-Finder outperformed all the other three tools on the synthetic examples, which contained degenerate motifs. MEME and DREME did not find the motifs in any case, while XXmotif found a similar result in 1 out of the 3 tests. The other 15 examples were taken from DNA and RNA high-throughput experiments~\cite{smeenk-27},\cite{harbison-28},\cite{hogan-29}. In 12 out of
these 15 datasets, mmHG-Finder found the motifs which were compatible with the known literature motifs as the most significant result. In comparison, DREME found the known consensus in 11 cases; XXmotif detected the literature motif in 9 cases while MEME detected the known motif in only 7 cases. In several datasets, such as for GCN4 and Pin4, mmHG-Finder identified the consensus motif while other tools returned repetitive sequences as their top results. The mmHG statistics avoids such spurious results as they typically do not correlate with the measurement driven ranking.

\begin{table}[!h]
\setlength{\tabcolsep}{1mm}
\caption{We evaluated the performance of mmHG-Finder in comparison to other state-of-the-art methods: MEME, DREME and XXmotif. Almost all input examples comprised ranked lists, except for p53 (comprising target and background sets). Since MEME, DREME, and XXmotif expect a target set as input, we converted the ranked lists into target sets by taking the top 100 sequences for MEME (restricted by MEME's limitation of 60,000 characters) and the top 20 \% sequences for the other tools. In the synthetic examples the entire ranked lists were taken as they are sufficiently small. Data and consensus motifs for p53 were taken from~\cite{smeenk-27}; for REB1, CBF1, UME6, TYE7, GCN4 from~\cite{harbison-28}; and for the RNA binding proteins from~\cite{hogan-29}. Selected results are shown below.}%
\centering
\begin{tabular}{|p{23mm} p{23mm} p{22mm} p{22mm} p{22mm}|}  \hline 
The protein and its consensus binding motif & mmHG-Finder & MEME & DREME & XXmotif \\ \hline
\textbf{Synthetic} TNWMNG W=[A/T], M=[A/C] & P\(\leq\)2.76e-14 \includegraphics[scale=0.08]{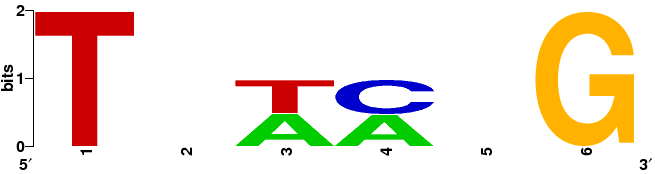}
& 7.0e+003 \includegraphics[scale=0.25]{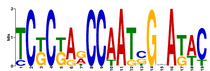}
 & Nothing found & 2.98e+00 \includegraphics[scale=0.3]{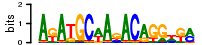} \\ \hline
\textbf{Synthetic} CTNNNAT & P\(\leq\)1.32e-28 \includegraphics[scale=0.1]{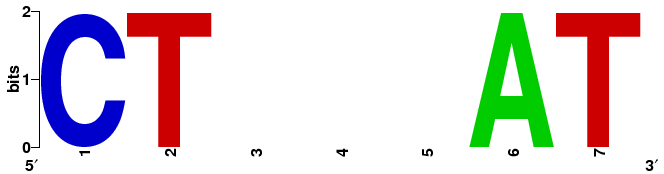}
& 7.1e+001 \includegraphics[scale=0.3]{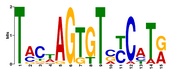} & Nothing found & 1.84e+01 \includegraphics[scale=0.3]{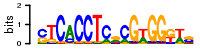} \\ \hline
\textbf{Synthetic} \mbox{MMMMMMMM} M=[A/C] & P\(\leq\)1.07e-39 \includegraphics[scale=0.1]{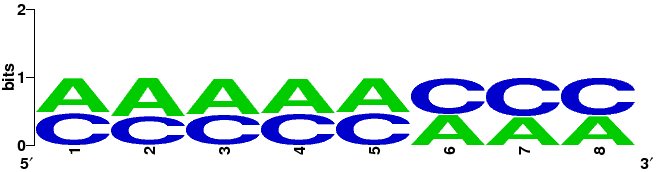}
& 1.8e+002 \includegraphics[scale=0.2]{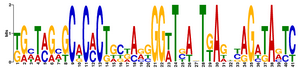} & Nothing found
& 1.58e+01 \includegraphics[scale=0.3]{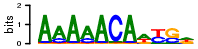} \\ \hline
\textbf{P53 (DNA)} \includegraphics[scale=0.2]{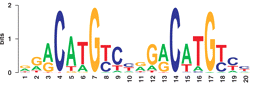} & P\(\leq\)1.09e-174 \includegraphics[scale=0.08]{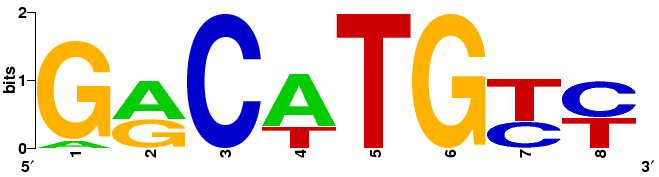}
& 1.8e-100 \includegraphics[scale=0.2]{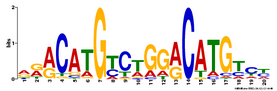} & \centering4.9e-133 \includegraphics[scale=0.3]{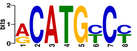}
& 1e-490 \includegraphics[scale=0.31]{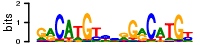} \\ \hline
\textbf{GCN4 (DNA)} {\bf\textcolor{red}{T}\textcolor{yellow}{G}\textcolor{green}{A}s\textcolor{red}{T}\textcolor{blue}{C}\textcolor{green}{a}}
& P\(\leq\)2.05e-44 \includegraphics[scale=0.08]{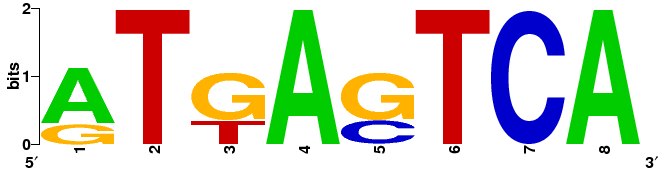}
& 1.3e-85 \includegraphics[scale=0.18]{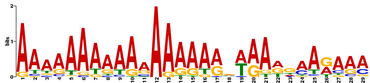} & \centering2.0e-32 \includegraphics[scale=0.3]{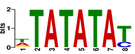}
& 4.00e-17 \includegraphics[scale=0.3]{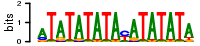} \\ \hline
\textbf{Puf5 (RNA)} \includegraphics[scale=0.45]{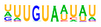} & P\(\leq\)7.93e-79 \includegraphics[scale=0.08]{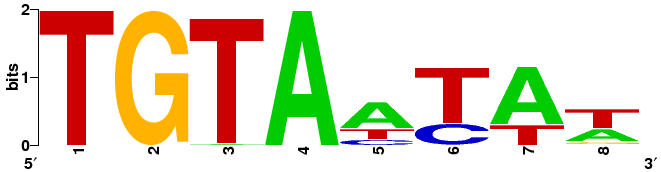}
& 3.6e-9 \includegraphics[scale=0.2]{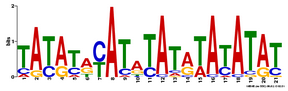} 3.1e-004 \includegraphics[scale=0.25]{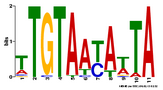} 
& \centering 6.8e-42\ \ \ \includegraphics[scale=0.36]{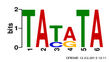}\ \ \ 3.1e-012 \includegraphics[scale=0.35]{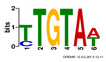}
& 9.76e-21 \includegraphics[scale=0.3]{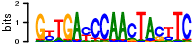} 1.61e-20 \includegraphics[scale=0.3]{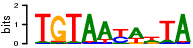} \\ \hline
\textbf{Pin4 (RNA)} \includegraphics[scale=0.45]{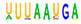} & P\(\leq\)8.18e-8 \includegraphics[scale=0.08]{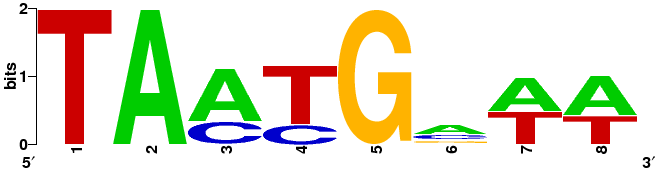}
& 1.3e+0 \includegraphics[scale=0.2]{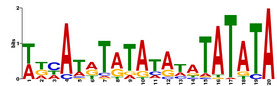}
& \centering3.1e-51\ \includegraphics[scale=0.37]{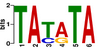}
& 1.65e-28 \includegraphics[scale=0.3]{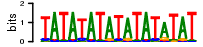}\\ \hline
 \end{tabular}%
 \label{tab:genedescription}%
\end{table}%
Computing p-value bounds for the synthetic examples (\textit{N}=500) took 7-17 seconds on a simple single-core laptop. The running time depends on both the number of elements \textit{N} as well as the mmHG score. The computation is optimized such that it is quicker for smaller mmHG scores. It took 33 minutes for \textit{N}=5000 where the mmHG score was \(3.3\cdot10^{-69}\), and 39 minutes for \textit{N}=4000 and 
mmHG score = \(5.9\cdot 10^{-31}\).

\section{Concluding Remarks}

Due to the size of the measure space, statistics over \(S_N\) is difficult to implement. We derive polynomially computable bounds for the tail distribution of the mutual enrichment at the top of two permutations uniformly and independently drawn over \(S_N\). We assess tightness using simulated data. We also demonstrate utility of the mmHG statistics in identifying motifs in experimental binding affinity data. For several representative datasets, including synthetically generated data, we note that our bound improves the p-value estimates by a factor of 50. The full characterization of the distribution of mmHG as a random variable over \(S_N\) remains an open question.

\subsubsection*{Acknowledgments.} We thank Israel Steinfeld for critical and inspiring discussions. We also thank the anonymous reviewers for their useful comments. LL was partially supported by Israel Ministry of Science and Technology and by ISEF Fellowship.

\bibliographystyle{spnlcs03}

\end{document}